\documentclass{emulateapj}

\newcommand{\Rs}{{\cal R}}

\slugcomment{Submitted to the Astronomical Journal}
\shorttitle{Deep NIR Imaging from the MUSYC Survey}
\shortauthors{Quadri et al.}

\begin{document}

\title{The Multiwavelength Survey by Yale-Chile (MUSYC): Deep Near-Infrared
Imaging and the Selection of Distant Galaxies}

\author{Ryan Quadri,\altaffilmark{1}
  Danilo Marchesini,\altaffilmark{1,2}
  Pieter van Dokkum,\altaffilmark{1,2}
  Eric Gawiser,\altaffilmark{1,2,3,4}
  Marijn Franx,\altaffilmark{5}
  Paulina Lira,\altaffilmark{3}
  Gregory Rudnick,\altaffilmark{6}
  C. Megan Urry,\altaffilmark{2}
  Jos\'{e} Maza,\altaffilmark{3}
  Mariska Kriek,\altaffilmark{5}
  L. Felipe Barrientos,\altaffilmark{7}
  Guillermo Blanc,\altaffilmark{8}
  Francisco J. Castander,\altaffilmark{9}
  Daniel Christlein,\altaffilmark{1,2,3}
  Paolo S. Coppi,\altaffilmark{1,2}
  Patrick B. Hall,\altaffilmark{10}
  David Herrera,\altaffilmark{1,2}
  Leopoldo Infante,\altaffilmark{7}
  Edward N. Taylor,\altaffilmark{5}
  Ezequiel Treister,\altaffilmark{11}
  Jon P. Willis\altaffilmark{12}}

\altaffiltext{1}{Department of Astronomy, Yale University, New Haven, CT}
\email{quadri@astro.yale.edu}
\altaffiltext{2}{Yale Center for Astronomy and Astrophysics, Yale
  University, New Haven, CT}
\altaffiltext{3}{Departamento de Astronom\'\i{}a, Universidad de
  Chile, Santiago, Chile}
\altaffiltext{4}{National Science Foundation Astronomy and
  Astrophysics Postdoctoral Fellow}
\altaffiltext{5}{Leiden Observatory, Universiteit Leiden, The
  Netherlands}
\altaffiltext{6}{National Optical Astronomical Observatory, Tucson,
  AZ}
\altaffiltext{7}{Departamento de Astronom\'\i{}a y Astrof\'\i{}sica,
  Pontifica Universidad Cat\'{o}lica de Chile, Santiago, Chile}
\altaffiltext{8}{Department of Astronomy, University of Texas at
  Austin, Austin, Texas}
\altaffiltext{9}{Instiut d'Estudis Espacials de Catalunya, Barcelona,
  Spain}
\altaffiltext{10}{Department of Physics and Astronomy, York University,
  Toronto, Canada}
\altaffiltext{11}{European Southern Observatory, Santiago, Chile}
\altaffiltext{12}{Department of Physics and Astronomy, University of
  Victoria, Victoria, Canada}

\begin{abstract}
  We present deep near-infrared $JHK$ imaging of four $10\arcmin
  \times 10\arcmin$ fields.  The observations were carried out as part
  of the Multiwavelength Survey by Yale-Chile (MUSYC) with ISPI on the
  CTIO 4m telescope.  The typical point source limiting depths are $J
  \sim 22.5$, $H \sim 21.5$, and $K \sim 21$ ($5\sigma$; Vega).  The
  effective seeing in the final images is $\sim 1\farcs0$.  We combine
  these data with MUSYC $UBVRIz$ imaging to create $K$-selected
  catalogs that are unique for their uniform size, depth, filter
  coverage, and image quality.  We investigate the rest-frame optical
  colors and photometric redshifts of galaxies that are selected using
  common color selection techniques, including distant red galaxies
  (DRGs), star-forming and passive BzKs, and the rest-frame
  UV-selected BM, BX, and Lyman break galaxies (LBGs).  These
  techniques are effective at isolating large samples of high redshift
  galaxies, but none provide complete or uniform samples across the
  targeted redshift ranges.  The DRG and BM/BX/LBG criteria identify
  populations of red and blue galaxies, respectively, as they were
  designed to do.  The star-forming BzKs have a very wide redshift
  distribution, a wide range of colors, and may include galaxies with
  very low specific star formation rates.  In comparison, the passive
  BzKs are fewer in number, have a different distribution of $K$
  magnitudes, and have a somewhat different redshift distribution.  By
  combining these color selection criteria, it appears possible to
  define a reasonably complete sample of galaxies to our flux limit
  over specific redshift ranges.  However, the redshift dependence of
  both the completeness and sampled range of rest-frame colors poses
  an ultimate limit to the usefulness of these techniques.
\end{abstract}


\section{Introduction}
\label{sec:introduction}

The issues surrounding galaxy formation and evolution provide
significant motivation for ongoing astrophysical research.  Although
several of the basic physical processes that drive the formation and
evolution of galaxies were identified some time ago, such as the
gravitational collapse of dark matter, the cooling and dissipative
collapse of baryons, the variation in star formation rates with time,
and galaxy mergers, constructing detailed models that reproduce
observed properties for a wide range of galaxy types and redshifts has
proven exceedingly difficult.  A precise determination of the
cosmological parameters has removed some of the uncertainties, but it
is largely accepted that comprehensive observations of galaxies at low
and high redshifts are necessary to refine our understanding of galaxy
evolution.

Recent years have seen dramatic progress in the observational study of
high redshift galaxies.  Much of this progress has been driven by deep
imaging of ``blank'' fields with multiple bandpasses.  These surveys
often rely on carefully-designed color selection techniques, making
use of only a few bands, to isolate samples of high redshift galaxies.
The most well known color selection criteria identify the Lyman Break
Galaxies (LBGs) at $z \sim 3$ \citep{steidel96}, using the $U_nGR$
bands.  This technique has been extended to select LBGs at higher
redshifts, and to select similar star-forming galaxies at $z \sim 2.3$
and $z \sim 1.7$ \citep[BX and BM galaxies,
respectively;][]{adelberger04}.  Selecting galaxies in optical bands
may miss many red high redshift galaxies, so NIR selection techniques
have also become important.  Extremely Red Objects (EROs) are at $z
\gtrsim 1$, and have been selected using a variety of criteria, such
as $R-K>5$ and $I-H>3$ \citep{mccarthy04}.  \citet{franx03} used
$J-K>2.3$ to select distant red galaxies (DRGs) at $z \sim 2-4$.
\citet{daddi04} proposed a technique involving the $BzK$ bands to
identify galaxies at $z > 1.4$.  \citet{yan04} used $(z-3.6\mu
\rm{m})_{\rm{AB}} > 3.25$ to isolate IRAC Extremely Red Objects
(IEROs).

The primary advantage of such selection techniques is that they rely
on only a few observed bands to isolate large samples of galaxies in a
(hopefully) well-defined redshift range.  However, diagnostic
information about individual galaxies or about the range of properties
spanned by a sample of galaxies requires more bands to trace the
detailed spectral energy distributions over a wide wavelength range.
Specific color selection techniques may also identify only a subset of
galaxies in the targeted redshift range (by design), offering a
limited view of the diversity of galaxy properties at that redshift.
Surveys with a large number of observed bands may be able to use
photometric redshifts to obtain a more complete sample of galaxies in
any given redshift range, and are better able to specify the intrinsic
properties of high redshift galaxies.

With these issues in mind, we have executed a deep optical/NIR survey
of four southern and equatorial fields as part of the Multiwavelength
Survey by Yale-Chile
(MUSYC).\footnote{http://www.astro.yale.edu/MUSYC}  Subsets of these
data have been used to study several characteristics of $K$-selected
galaxies at $z>2$, including the number density and colors of massive
galaxies \citep{vandokkum06}, the clustering properties
\citep{quadri06}, and the luminosity function \citep{marchesini06}.
Additionally, \citet{kriek06a,kriek06b,kriek06c} obtained NIR
spectroscopy of the brightest MUSYC galaxies, and \citet{webb06} used
\emph{Spitzer} MIPS imaging to infer the star formation rate of MUSYC
DRGs.

The purpose of this paper is two-fold.  First, we present the MUSYC
deep NIR imaging.  We describe the observations, data reduction, and
the $K$-selected catalogs with full $UBVRIzJHK$ photometry.  Second,
we use these catalogs and high quality photometric redshifts to
investigate the properties of galaxies that are selected with common
color selection criteria.  We use Vega magnitudes unless noted
otherwise.  Throughout we use $H_0 = 70~\rm{km^{-1}s^{-1}Mpc^{-1}}$,
$\Omega_m = 0.3$, and $\Omega_{\Lambda}=0.7$.

\section{The Multiwavelength Survey by Yale-Chile}

The MUSYC survey consists of several components: optical imaging of
four $30\arcmin \times 30\arcmin$ fields, NIR imaging over the same
area, deeper NIR imaging over four $10\arcmin \times 10\arcmin$
subfields, and spectroscopic followup.  Some of the MUSYC fields also
benefit from imaging by \emph{Chandra} and \emph{XMM} in the x-ray,
\emph{HST} in the optical, and \emph{Spitzer} in the infrared.

The survey design is described by \citet{gawiser06a}.  The four
$30\arcmin \times 30\arcmin$ fields were selected to have low Galactic
reddening, HI column density \citep{burstein78}, and $100\mu m$ dust
emission \citep{schlegel98}.  They were also chosen to have high
Galactic latitude to reduce the number of stars, to cover a wide range
in right ascension to enable flexible scheduling, and to be accessible
from Chilean observatories.  These fields were observed with $UBVRIz$
filters for complete optical coverage.  A narrowband $5000$\AA\ filter
was also used to identify Lyman $\alpha$ emitters at $z \simeq 3$
\citep{gawiser06b}.  \citet{gawiser06a} presents images and
optically-selected catalogs of one of the MUSYC fields, EHDF-S.  The
full optical data for the remaining fields will be described elsewhere
(E. Gawiser et al. 2007, in preparation).

The four $30\arcmin \times 30\arcmin$ fields were also observed in at
least one of the $J$, $H$, and $K$ bands.  We refer to this as the
``wide'' portion of the MUSYC NIR imaging.  These data were collected
and processed in a similar way to the data described in this paper
(E. Taylor et al. 2007, in preparation; G. Blanc 2007 et al. 2007, in
preparation).  The typical depths are $J \sim 22.$ and $K \sim 20$.

Four $10\arcmin \times 10\arcmin$ subfields were observed to greater
depth in all of $JHK$.  These data, which are referred to as the
``deep'' NIR MUSYC imaging, are presented in this paper.  Two of these
subfields, HDFS1 and HDFS2, are adjacent and lie within the larger
$30\arcmin \times 30\arcmin$ MUSYC Extended Hubble Deep Field-South
(EHDF-S).  The other two subfields lie within the larger MUSYC 1030 and
1255 fields.  We did not perform deep NIR imaging in the fourth large
MUSYC field, ECDF-S, because very deep imaging in the central region
of this field has been made available by the Great Observatories
Origins Deep Survey (GOODS) team \citep{giavalisco04}.  The exact
locations of the deep MUSYC fields within the larger $30\arcmin \times
30\arcmin$ fields were chosen to avoid bright stars.  Information
about these fields is given in Table~\ref{tbl:field_chars}.

\section{Observations}
\label{sec:observations}
The MUSYC deep NIR observations were performed with the Infrared
Sideport Imager (ISPI) \citep{probst03,bliek04} on the CTIO Blanco 4m
Telescope.  The detector is a $2048 \times 2048$ HgCdTe HAWAII-2
array, with a $\sim 0\farcs305$ pixel scale and a $10\farcm5
\times 10\arcmin.5$ field of view.  Observation were performed
over the course of 9 observing runs from January 2003 to April 2006.

The mean airmass of observation varied between 1.21 and 1.43 for
different field/filter combinations, and exposures were rarely taken
at airmass $>1.6$.  Standard stars from \citet{persson98} were
observed 2-4 times per night except when the conditions were poor.
The range of airmass values for standard star observations was similar
to that of science observations.

The background emission in the NIR is bright, non-uniform across the
field, and can vary on short timescales.  Accurate background
subtraction requires that the telescope be dithered between exposures
(see \S \ref{sec:data_reduction}).  Because the brightest objects in an
exposure can leave residual images in subsequent exposures, using a
non-regular dither pattern -- in which the telescope does not
repeatedly trace the same sequence of dither positions -- facilitates
removal of artifacts during the reduction process.  We used an
algorithm to generate semi-random dither patterns in which the
distance between subsequent dither positions is maximized.  The size
of the dither box is $45\arcsec$, which is sufficiently large to
obtain good background subtraction in the regions around all but the
brightest/most extended sources in our fields without significantly
reducing the area with the highest exposure times.

The typical exposure times at each dither position were $1 \times
100\rm{s}$ (coadds $\times$ individual exposure time) in $J$, $4
\times 20\rm{s}$ in $H$, and $4 \times 15\rm{s}$ in $K$.  The total
exposure times for each field/filter combination, after discarding
images with poor quality (\S \ref{sec:data_reduction}), are given in
Table~\ref{tbl:field_chars}.

\begin{figure}
  \epsscale{1.1}
  \plotone{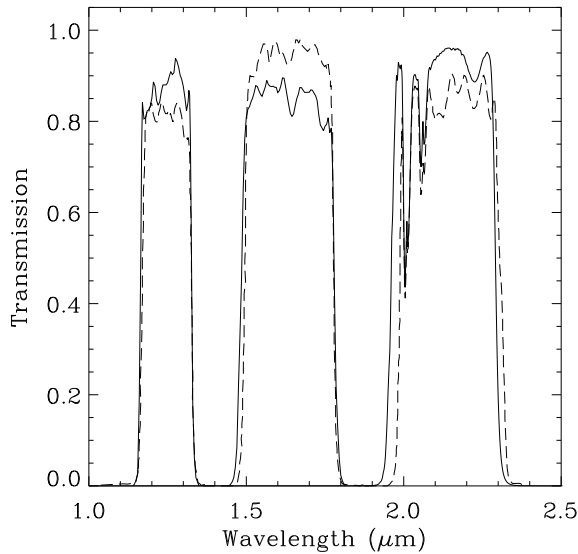}
  \caption{The filter transmission curves, including atmospheric
  transmission, used for the MUSYC NIR imaging.  The solid curves are
  for HDFS1 and 1030 and the dashed curves are for HDFS2 and 1255.}
  \label{fig:filt_curves}
\end{figure}

The set of filters used with ISPI changed in April 2004.  The fields
HDFS1 and 1030 were completed using the original $JHK'$ filters,
whereas HDFS2 and 1255 were observed using the newer $JHK_s$ filters.
Both sets of filter transmission curves are shown in
Figure~\ref{fig:filt_curves}.  Conversions between the Vega and AB
magnitude systems were calculated using the SED of Vega, and are
given, along with the effective wavelengths, in
Table~\ref{tbl:AB_conversions}.  Note that the shift in effective
wavelength between the $K'$ and $K_s$ filters is small, and much less
than the filter widths.  In the discussion that follows we do not
distinguish between the two filter sets.

\section{Data Reduction and Image Properties}
\label{sec:data_reduction}

\subsection{Data Reduction}
The data were reduced using a combination of standard
IRAF\footnote{IRAF is distributed by the National Optical Astronomy
  Observatories, which are operated by the Association of Universities
  for Research in Astronomy, Inc., under cooperative agreement with
  the National Science Foundation.} tasks, modified IRAF tasks, and
custom tasks.  The core of our reduction procedure was the external
IRAF package
XDIMSUM.\footnote{\url{http://iraf.noao.edu/iraf/ftp/iraf/extern-v212/xdimsum020806}}
The basic methods are similar to those described in detail by
\citet{labbe03}, and are outlined below.

Near-infrared cameras have a non-negligible dark current, so dark
images with the appropriate exposure time and number of coadds are
subtracted from the science images.  We use domeflats to perform the
flat field correction.  To remove the background emission from the
domeflat image, and thus to isolate the uniform illumination of the
dome screen by the lamp, we follow the standard procedure of
subtracting a ``lamp off'' image from a ``lamp on'' image.  During
the observations, special care was taken to keep the count level of
the ``lamp on'' images at a reasonable level, well below where
non-linearity becomes significant on the ISPI camera.  Domeflats were
taken nightly or semi-nightly; they were generally stable from night
to night, but showed significant variations between observing runs.

The background emission in each science frame is subtracted, and the
resulting images combined, in two passes.  During the first pass, an
image of the background emission is created for each science image
using a running median of the dithered sequence of science images;
e.g.~the background for image 10 is the median combination of images
6-9 and 11-14.  The positions of several stars are used to determine
the relative shifts between background-subtracted images.  The images
are shifted to a common reference using sub-pixel interpolation and
are combined.  Objects in the combined image are detected using a
simple thresholding algorithm.  These masks are shifted back into the
frame of individual science exposures, and the background-subtraction
process is repeated during the second pass; this time objects are
masked out during the calculation of the running median in order to
improve the background subtraction.

We take several steps to improve the quality of the final images.  We
create a mask of bad pixels in each image.  An initial list of bad
pixels is created using the flat-field images.  We then inspect each
background-subtracted image individually; images with severe artifacts
(e.g.~disturbed point spread functions) are discarded, while others
with localized artifacts (e.g.~satellite trails) are masked using a
custom procedure.  Additional bad pixels in each image are identified
using a cosmic ray detection procedure or are removed with a
sigma-clipping algorithm during image combining.  The ISPI array can
retain memory of previous exposures in the form of persistence images
of bright objects.  We create a second object mask at the end of the
first pass reduction, in which only the cores of the brightest objects
are masked; these are used to mask the pixels that contained bright
objects during the previous exposure.  Finally, we optimize the
signal-to-noise in the seeing disk by using a weighted average during
the final image combining step.  The weights are calculated using
\begin{equation}
w_i = \frac{1}{(scale_i*rms_i*FWHM_i)^2,}
\end{equation}
where $scale_i$ is a constant used to scale the signal in image $i$ to
the common level, and $rms_i$ is the pixel-to-pixel rms measured in a
blank region of the unscaled image.  The values $scale_i$ and $FWHM_i$
are determined for each image using a set of bright non-saturated
stars, which are chosen in locations away from dense regions of bad
pixels.

The final products of our reduction procedures are a combined image,
an exposure time map, and an rms map which gives an estimate of the
noise level at each pixel.

\subsection{Astrometric Correction and Optical Images}

The observations, reduction, and characteristics of the first optical
MUSYC data are described in \citet{gawiser06a}.  Subsequent data were
reduced and analyzed using similar methods, and will be described by
E. Gawiser et al.~(2007, in preparation).  Most of the optical imaging
was obtained using the 8 CCD MOSAIC II camera on the Blanco 4m
telescope at CTIO.  Each image was re-sampled to provide a uniform
pixel scale and tangent plane projection using stars with known
positions from the USNO-B catalog.  The final rms astrometric errors
are estimated to be less than $0\farcs2$ across the entire field.

We use the IRAF tasks GEOMAP and GEOTRAN to re-sample the NIR images
so that they follow the same logical and world coordinate system
(i.e. $x$, $y$ pixel coordinates and $RA$, $Dec$) as a set of trimmed
optical MUSYC images.  This process flips the NIR images around the
x-axis so that North is up and East is left, in accordance with
standard practice.  The pixel scale is changed slightly, from
$0\farcs305$ to $0\farcs267$.  We used a $6th$ order fit in x and y,
allowing for cross-terms, to adequately remove the distortions present
in the ISPI instrument; a similar high-order fit was also found to be
necessary by ISPI instrument scientists \footnote{see
  \url{http://www.ctio.noao.edu/instruments/ir\_instruments/ispi/}}.
We verified that the pixel resampling does not introduce systematic
errors in flux for a set of objects distributed across the images.
The rms error in stellar positions between the NIR and optical images
is $\sim 0\farcs06$, much smaller than the typical aperture size used
for photometry.  We also compared the positions of stars in our
$K$-band images to those in the USNO-B catalog directly.  Many of the
objects are in fact extended, and some others are saturated in our
images; after removing these objects the rms is $0\farcs15-0\farcs2$,
consistent with the uncertainties of individual USNO-B stars.

\begin{figure*}
  \plotone{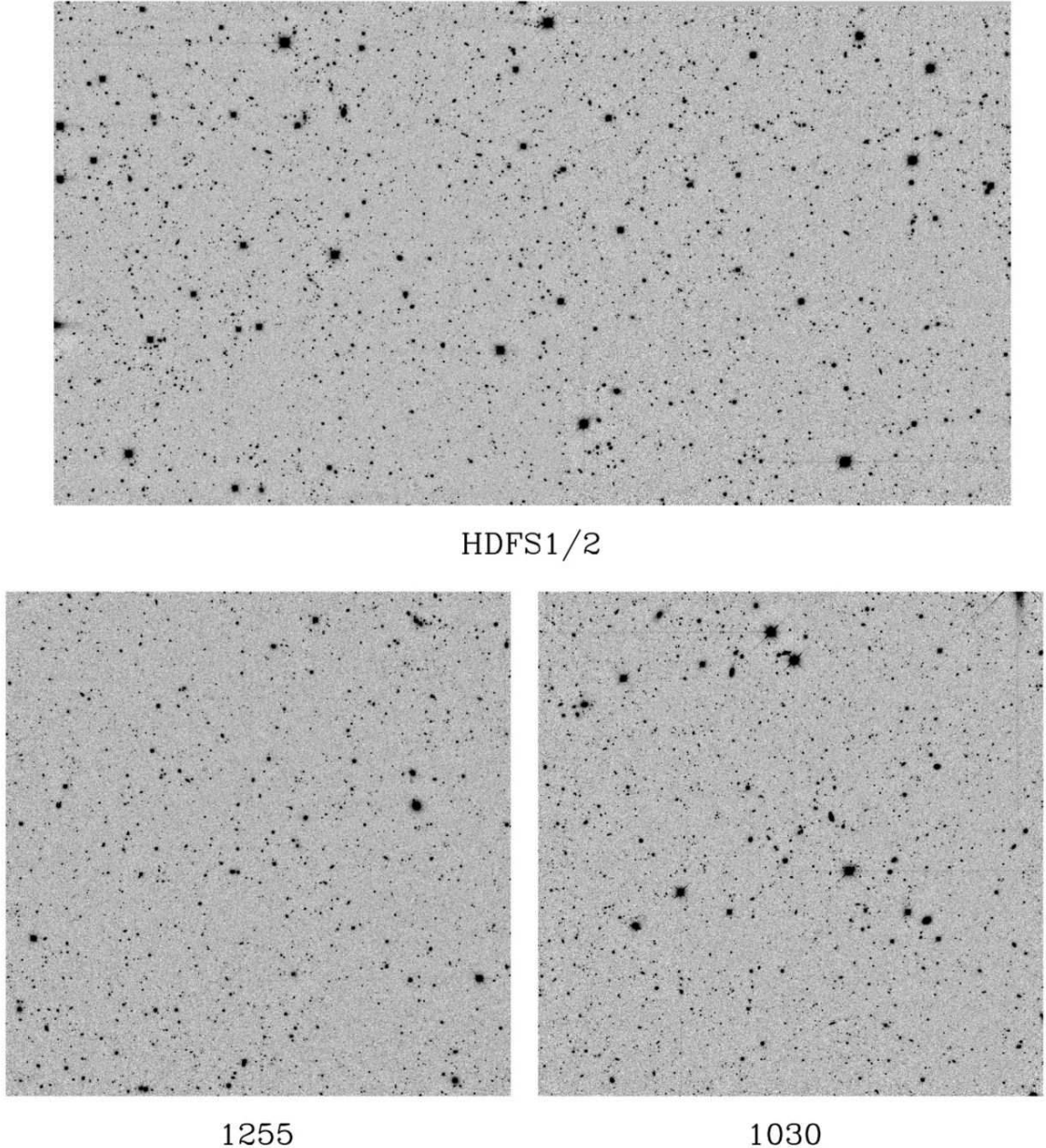}
  \caption{The deep $K$-band images from the MUSYC.  At top, the
    individual HDFS1 and HDFS2 images have been combined to create a
    single $19\farcm5 \times 10\farcm3$ image.  The other fields are
    $\sim 10\farcm3$ on a side.}
  \label{fig:images}
\end{figure*}

\begin{figure*}
  \plotone{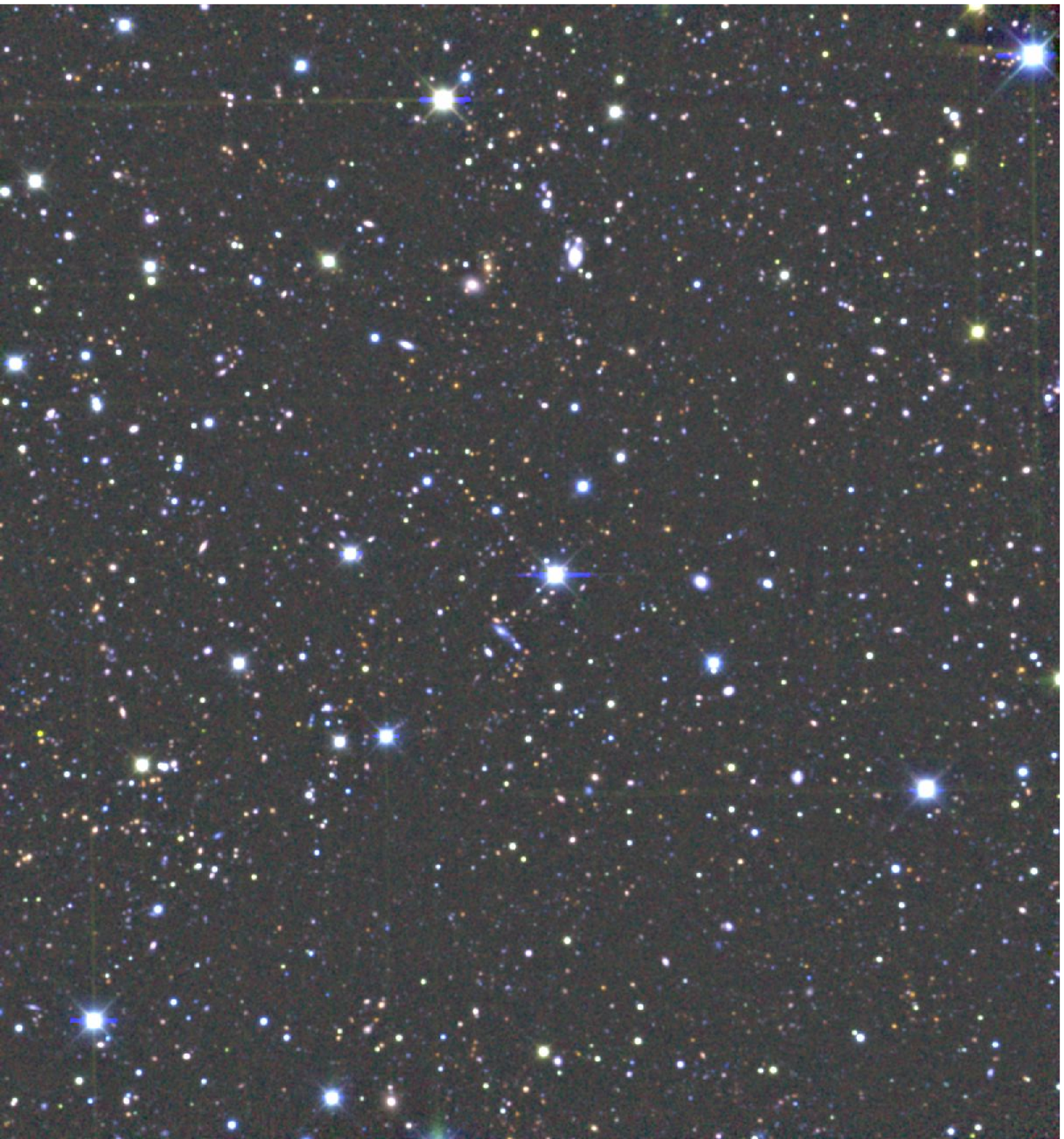}
  \caption{An $RJK$ composite of HDFS1}.
  \label{fig:color_image}
\end{figure*}

The final $K$-band images are shown in Figure~\ref{fig:images}.  The
stretch has been adjusted to emphasize faint sources and the
uniformity of the background.  Figure~\ref{fig:color_image} shows a
color composite image of HDFS1, constructed using the $RJK$ bands.
The effective seeing in the final images is taken as the median of the
FWHM for a set of $\sim 5-10$ stars, and is given in
Table~\ref{tbl:field_chars}.

\subsection{Photometric Calibration and Verification}

Observations of standard stars from \citet{persson98} were performed
nightly except when the conditions were very poor.  Each standard
star observation used a large 5-point dither pattern, and the
telescope was de-focused to keep the peak count level in the linear
regime of ISPI.  Following \citet{persson98}, the instrumental
magnitude of the star was measured in a $10\arcsec$ diameter aperture;
we verified that increasing the aperture size would change the
instrumental zeropoints by $<1\%$.  The zeropoint rms from a single
dither sequence of a standard star is typically $0.01-0.03$, setting
an approximate scale for the flatness of the images.  We then
calibrated a set of secondary standard stars in the science field with
the same size aperture, after applying a small correction for the
different airmass of standard star and science observations.  It was
not possible to determine accurate airmass coefficients from our
observations, so we used the coefficients given by \citet{frogel98},
which were nonetheless found to be consistent with our observations.
Following usual practice in the NIR, we did not account for color
terms in the photometric calibration.

There was at least one night for each field/filter combination where
the observing conditions were sufficient to determine an accurate
calibration of the secondary standards.  For field/filter combinations
where there was more than one night of satisfactory calibrations, the
agreement between nights is typically $0.01-0.02$.  While it is
difficult to place constraints on possible systematic uncertainties,
the internal precision of our final zeropoints is $\lesssim 0.03$ in
every case.  The zeropoints are given in Table \ref{tbl:image_chars}.

The quality of our photometric calibrations was verified in several
ways.  There is a $\sim 10~\textrm{arcmin}^2$ overlap region between
two of our fields, HDFS1 and HDFS2, where the exposure time is at
least half of the total.  As these fields were observed on different
runs, using different NIR filters, and the photometric zeropoints were
determined independently, comparing the total magnitudes
(\S~\ref{sec:photometry}) of objects in this region from the two
fields allows a check of the internal consistency of our photometric
calibrations.  We find that the median offset is less than 0.02 for
bright objects in each of $J$, $H$, and $K$.  It should be noted that
these objects lie near (opposite) edges of the ISPI detector in the
two fields, where the flat-fielding errors may be larger; the good
agreement between the two fields is encouraging.  We do not attempt to
adjust the zeropoints to obtain better agreement.

The MUSYC HDFS1 field contains the $\sim 4.5~\textrm{arcmin}^2$ FIRES
HDF-S field \citep{labbe03}, allowing for an additional check of our
photometry.  The FWHM of the FIRES image is $\sim 0\farcs45$.  Rather
than using the total magnitudes from the FIRES catalog, we convolved
the public FIRES image to match the MUSYC point spread function (PSF),
and directly compared the aperture photometry for a set of compact
objects.  We found the agreement to be better than $0.02$ magnitudes
in each of $JHK$.

A final check comes from comparing our total magnitudes to the
aperture-corrected magnitudes from the public 2MASS point-source
catalog \citep{skrutskie06}.  We selected a set of matching objects from
the three MUSYC deep NIR fields, rejecting objects with low S/N in the
2MASS photometry, that were affected by non-linearity/saturation in
the MUSYC photometry, or were blended with other objects.  The mean
offset ranges between $0.01-0.04$ for HDFS1 and HDFS2, generally
consistent with the errors.  However, adopting zeropoints based on the
2MASS photometry would decrease the level of agreement for objects in
the overlapping region of HDFS1 and HDFS2 for two out of three NIR
filters, and would decrease the agreement between HDFS1 and the FIRES
HDF-S photometry for all three filters.  For these two fields, we
conclude that our flux calibration and the quoted zeropoint
uncertainties are reliable, and are better than could be obtained by
calibrating directly from the 2MASS catalogs.

However, the agreement between 2MASS photometry and MUSYC photometry
is worse for the 1030 and 1255 fields.  The mean difference between
2MASS and MUSYC 1030 photometry is a remarkably consistent $\sim 0.050
\pm 0.015$ in each of $JHK$, where the uncertainties are given as the
standard deviation of the mean.  The sense of the disagreement is that
the stars are brighter in the 2MASS catalog than in the MUSYC catalog.
The disagreement is of similar size, but in the opposite direction,
for 1255.  This is worrisome, as the level of disagreement for these
fields in all three filters is worse than in any of the filters for
HDFS1 and HDFS2, and moreover the differences are systematic across
the filters.  We have verified that this disagreement is not caused by
obvious mistakes in our photometric calibrations.  The systematic
nature of the offsets \emph{may} indicate that they are caused by
aperture corrections in 2MASS, rather than zeropoint errors.  Because
our calibrations appear to work very well for HDFS1 and HDFS2, we do
not adjust our zeropoints using 2MASS photometry; but without other
data to compare with, we cannot be certain that our $\lesssim 0.03$
magnitude zeropoint uncertainties for these fields are reliable.

\subsection{Noise Properties and Limiting Depths}

The flux uncertainty within an aperture has a contribution from the
photon statistics from astronomical objects, as well as a contribution
from background noise, which is due to sky emission, read noise, etc.
The standard method of determining the contribution of background
noise, by e.g.~SExtractor \citep{bertin96} or the APPHOT package of
IRAF, is to scale the rms flux value of background pixels by the
square root of the number of pixels within the aperture,
$\sqrt{N_{pix}}$.  This scaling is appropriate if adjacent background
pixels are uncorrelated, however correlations will be introduced by
imperfect subtraction of background emission, extended wings from
bright objects, undetected sources, re-pixelization during the
reduction procedure, and artifacts in the images.  In the limiting
case of perfect correlations between the pixels within an aperture,
the background noise $\sigma_{back}$ will scale as $N_{pix}$.  Thus it
might be expected that the true scaling will be $\sigma_{back} \propto
N_{pix}^\beta$, where $0.5<\beta<1$.

\begin{figure}
  \epsscale{1.1}
  \plotone{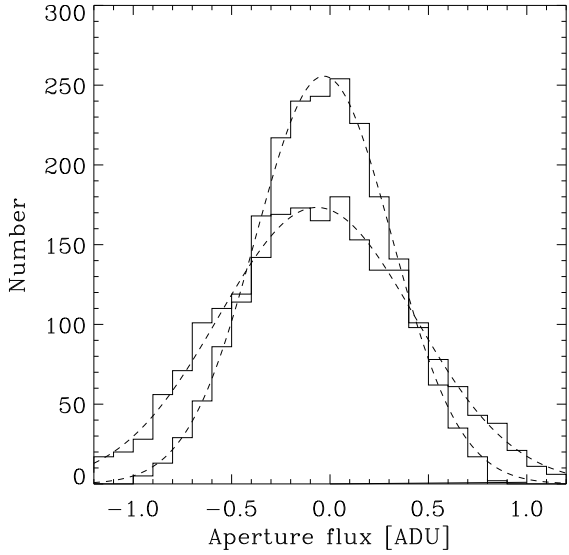}
  \caption{Histograms of background fluctuations in $1\farcs5$ and
    $2\arcsec$ diameter apertures in the HDFS1 $K$-band image.  The
    larger aperture has the broader distribution of fluxes.  The
    dashed curves are the best-fitting Gaussians.}
  \label{fig:noise_hist}
\end{figure}

\begin{figure}
  \epsscale{1.1}
  \plotone{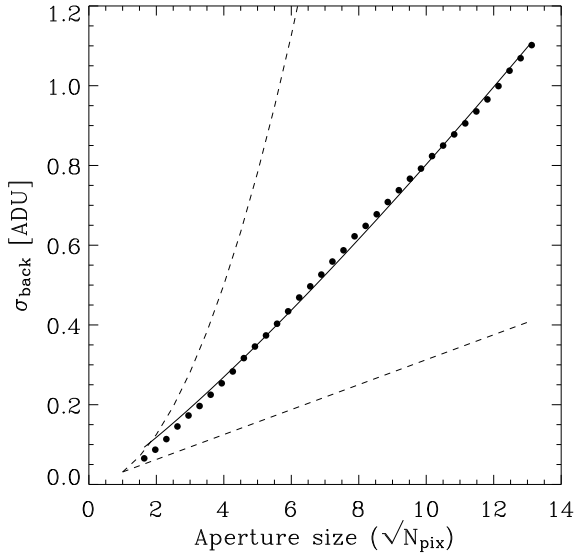}
  \caption{The rms value of background fluctuations within an aperture
    as a function of the aperture size for HDFS1 $K$.  The filled
    circles illustrate the measured values, while the solid curve is a
    power law fit.  The dashed curves show the expected scaling from
    the measured pixel-to-pixel rms relation in the case of no pixel
    correlations (lower) and perfect correlation of all pixels within
    each aperture (upper).}
  \label{fig:noise_apsize}
\end{figure}

We follow \citet{labbe03} in characterizing the noise properties of
our images by summing the counts in apertures distributed randomly
over empty regions of each image.  The rationale and method are
discussed more fully in \citet{gawiser06a}.
Figure~\ref{fig:noise_hist} shows the distribution of fluxes in
$1\farcs5$ and $2\arcsec$ diameter apertures.  These distributions are
well-approximated by Gaussians.  Figure~\ref{fig:noise_apsize} shows
how the width of the best-fitting Gaussian changes with aperture size,
along with predictions from the $\sqrt{N_{pix}}$ and $N_{pix}$
scalings that are described above.  It is apparent that neglecting
correlations between pixels would cause a significant underestimation
of background fluctuations.  A single power law provides a good fit to
the relationship between $N_{pix}$ and $\sigma_{back}$ for the
aperture sizes of interest; the typical power law index is $\beta \sim
0.6$.

We estimate the flux uncertainties for each object in each band as
\begin{equation}
  \sigma^2 = \sigma_{back}^2 + \frac{F}{GAIN},
\end{equation}
where $F$ is the flux in ADU, $GAIN$ is the total effective gain, and
$\sigma_{back}$ is estimated for an aperture with the appropriate
size.  In the case of the elliptical Kron apertures (see
\S~\ref{sec:photometry}) we estimate $\sigma_{back}$ for a circular
aperture with the same area.

The characteristics of the final images can be found in Table
\ref{tbl:image_chars}.  This table includes an estimate of the $5
\sigma$ point source limiting depths.  These values are calculated
using the background fluctuations in the color aperture (see \S
\ref{sec:phot_colors}), and with a $0.75$ magnitude aperture
correction applied to account for the $\sim 50\%$ point-source flux
that falls outside this aperture.  An alternate analysis of the
limiting depths can be found in \S~\ref{sec:completeness}.

\section{Source Detection and Photometry}
\label{sec:photometry}

\subsection{Source Detection}
We used the SExtractor v2.4.3 software \citep{bertin96} to detect
objects in the $K$-band images.  The two parameters that affect the
sensitivity of source detection are DETECT\_THRESH, which was used to
specify the detection threshold in units of the rms of background
pixels, and DETECT\_MINAREA, which specifies the number of adjacent
pixels that must meet this threshold.  Additionally, SExtractor
optionally filters the detection image with a convolution kernel prior
to detection in order to enhance the detection of faint objects.
Rather than use detailed simulations to find the optimum set of
parameters for SExtractor, we made use of the ultradeep K-band image
of HDFS from the public FIRES survey \citep{labbe03}.  The difference
in depth between the MUSYC HDFS1 $K$-band image and the FIRES HDF-S
image ($K \sim 21.0$ and $K \sim 24.3$, respectively) allows for a
clear determination of which objects detected by SExtractor in HDFS1
are real and which are noise peaks.  We increased the DETECT\_THRESH
parameter slightly from the default value in order to eliminate all
spurious sources in the FIRES region, but without significantly affecting
the completeness of our catalogs.  The final detection parameters were
DETECT\_MINAREA=5 pixels and $\rm{DETECT\_THRESH} = 1.6$, and we used
a Gaussian filter corresponding to the PSF of the detection image.  We
detected objects in rms-normalized $K$-band images, which were created
by dividing the $K$-band science images by the rms maps (see
\S\ref{sec:data_reduction}).  Detection in normalized images is
important because the dithering procedure used during observations
means that the outskirts of the science images are noisier, which
would otherwise result in a large number of spurious sources.

\subsection{PSF matching}
\label{sec:psf_matching}

In order to measure accurate colors, all optical and NIR images were
PSF matched using Gaussian convolution kernels.  This is necessary to
ensure that the aperture used for color measurements (\S
\ref{sec:phot_colors}) contains a constant fraction of the total light
for every object in every band.  The systematic effects on colors that
are introduced by imperfect PSF matching can be estimated by dividing
the stellar growth curves, which were created for each image using a
set of bright non-saturated stars evenly distributed over the images.
We optimized the width of the convolution kernels by minimizing the
difference in growth curves; we found that simple Gaussians are
capable of reducing the systematic errors to the $\sim 1-2\%$ level
for apertures as large or larger than the color apertures.  The
exception is the $z$-band, which has a PSF shape that is sufficiently
different from the other bands that a Gaussian kernel yields errors
that are $\sim 2-3\%$.

Most of the images have FWHM from $\sim 0\farcs8-1\farcs0$.  However the
$U$ and $B$ bands are slightly worse, with values as high as
$1\farcs4$.  Degrading all the images of a given field to the broadest
PSF would reduce the S/N significantly in every band.  For the images
with narrow PSFs, we smooth to $\sim 1\arcsec$ in order to measure
uniform colors.  Images with broader PSFs are treated differently.  We
measure e.g.~the $U-K$ color using a version of the $K$ image
that is degraded to the same PSF as $U$, and add this color to the
$K$-band flux in the color aperture of the $\sim 1\arcsec$ $K$ image.
It is important to note that this procedure may not be appropriate for
extended objects with strong color gradients.  This is because the
$U-K$ color is measured for a larger region of the galaxy than is
e.g.~the $J-K$ color, with the effect that the inferred $U-J$ color
will be different than what could be measured in any aperture.  This
effect should not be a significant concern here, because distant
galaxies are at best marginally resolved in our $\sim 1\arcsec$ seeing
and observed color gradients are minimal.  Indeed, we have verified
that any errors introduced by this procedure are at no more than the
percent level by directly measuring the $U-J$ color (using a smoothed
version of the $J$ image) and comparing it to what is inferred from
the $U-K$ and $J-K$ colors.

\subsection{Photometry and Colors}
\label{sec:phot_colors}

We use SExtractor in dual-image mode to detect objects in $K$ and to
perform photometry in all bands.  In this subsection we describe our
methods to determine ``total'' flux in $K$, and to measure high
signal-to-noise (S/N) colors.  The total flux in any other band can be
calculated directly from these quantities.

We estimate the total $K$-band flux of an object from the SExtractor
AUTO photometry, which uses a flexible elliptical Kron-like aperture.
Although this aperture contains most of the flux for bright objects,
in practice the aperture can become quite small for faint
sources--even for the same light profile--so a larger fraction of
light may be missed.  Therefore we convert the AUTO flux to total flux
by applying an aperture correction.  The aperture correction is
calculated using the median stellar growth curve of a set of bright
stars, for a circular aperture with the same area as the AUTO
aperture.  We note that this procedure may introduce a mild bias in
the total flux of faint, extended sources, because their light
profiles may not follow that of a point source outside the AUTO
aperture; however these sources will have highly uncertain flux
measurements no matter what procedure is used.  The aperture
correction for the AUTO aperture can reach $\sim 0.5$ magnitudes for
some of the faintest sources.

Because accurate color measurements are necessary for photometric
redshift calculations and modeling of stellar populations, we wish to
optimize the S/N.  Since the noise is a strong function of aperture
size (Fig.~\ref{fig:noise_apsize}), we measure the colors in smaller
apertures than the AUTO aperture that is used to estimate the total
flux.  For a point source with a Gaussian PSF, and for uncorrelated
background noise, the aperture with optimal S/N has diameter $1.35
\times FWHM$ \citep{gawiser06a}.  In realistic situations there are
competing effects which change the size of the optimal aperture.
Relative to the idealized case, the broader wings of our PSFs would
suggest a larger aperture to obtain the same signal, while the noise
correlations suggest a smaller aperture to keep the same level of
noise.  The optimal S/N aperture for a point source can be found by
dividing the stellar growth curve by the $\sigma_{back}$ curve shown
in Figure~\ref{fig:noise_apsize}.  We find that the optimal aperture
typically has diameter $\sim 1.1-1.4$ times the FWHM, depending on the
filter \citep[see also][]{gawiser06a}.  Using a very small aperture
for color determinations presents several problems.  Accurate colors
require that a similar fraction of the flux from an object is
contained within the color aperture in each filter, which is easier to
achieve with larger apertures.  This is because a larger aperture
contains a larger fraction of the flux, leading to a smaller relative
difference in the fraction of flux.  Secondly, one of the primary
scientific goals of the deep NIR MUSYC survey is the study of high
redshift galaxies, which may not appear as pure point-sources in our
$\sim 1\arcsec$ images.  Larger apertures also reduce the effects of
the small variations in PSF across the images, and the residual
geometric distortions at the edges of the images.  We choose, as a
compromise, apertures with diameter $\sim 1.5$ times the stellar FWHM.
These apertures contain $\sim 50\%$ of the light from a point source.
We verified that the S/N in these apertures is $\gtrsim 95\%$ of the
S/N in the optimal aperture for every image.

\subsection{Catalog format}

The photometry in the $K$-selected catalogs is presented in units of
flux, normalized so that the zeropoint is 25 on the AB system.  The
use of flux, rather than magnitudes, avoids the problem of converting
the measured flux uncertainties into magnitude uncertainties, the
problem of asymmetric magnitude uncertainties for low S/N objects, and
the loss of information for objects that have negative measured
fluxes.  The photometry is not corrected for Galactic extinction.

Object detection and measurement of geometrical parameters are
performed by SExtractor in the rms-normalized $K$ image.  The adjacent
HDFS1 and HDFS2 fields were treated separately, but a list of objects
that appears in both catalogs is available.  Version 3.1 of the
$K$-selected catalogs are given in the following format:\linebreak

\emph{Column 1:}  SExtractor ID number, starting with 1

\emph{Columns 2-3:}  x and y barycenters

\emph{Columns 4-5:}  RA and DEC (decimal degrees; J2000)

\emph{Column 6:}  internal MUSYC field code (4=HDFS1, 5=HDFS2, 6=1030, 7=1255)

\emph{Columns 7-24:} Flux density and uncertainty in the color aperture, in
the order $UBVRIzJHK$

\emph{Columns 25-26:} $K$-band flux density and uncertainty in the ``total''
aperture

\emph{Columns 27-35:} Exposure time weight in bands $UBVRIzJHK$, normalized
to the weight of the median object

\emph{Columns 36-37:} Diameters of color and AUTO apertures.  The diameter of
the AUTO aperture is taken as the geometric mean of the major and
minor axes (arcsec)

\emph{Columns 38:} SExtractor blending flag 2 -- object was originally
blended with another object (1=blended, 0=unblended)

\emph{Columns 39:} SExtractor blending flag 1 -- object was blended with
another object strongly enough to significantly bias AUTO photometry
(1=blended, 0=unblended)

\emph{Column 40:}  Half-light radius (arcsec)

\emph{Column 41:}  Ellipticity

\emph{Column 42:}  Position angle (degrees), measured counter-clockwise from North

\emph{Column 43:}  Aperture correction to convert AUTO flux to total flux

\emph{Column 44:}  SExtractor CLASS\_STAR parameter

\emph{Column 45:}  Maximum SExtractor flag, for bands $UBVRIzJH$

\emph{Column 46:}  SExtractor flag, $K$-band

The catalogs are available in the electronic edition of the journal
article.

\subsection{Completeness}
\label{sec:completeness}

\begin{figure}
  \epsscale{1.1}
  \plotone{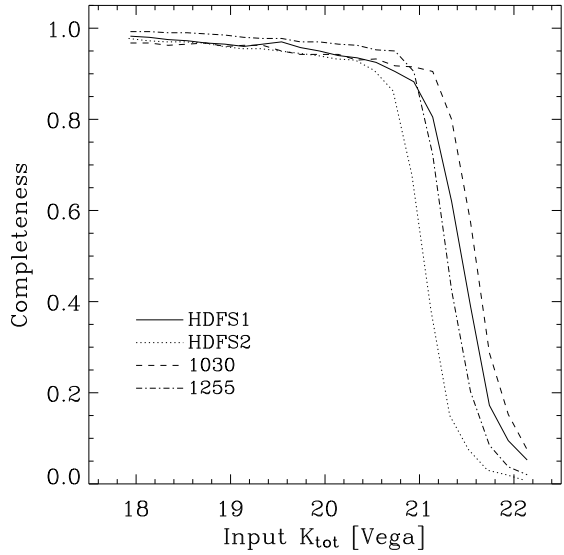}
  \caption{The $K$-band completeness curves.  Point sources were
    inserted at random locations in the central region of the four
    MUSYC fields, where the exposure time is $\gtrsim 95\%$ of the
    total.  The completeness is defined as the fraction of simulated
    sources that were recovered, as a function of total magnitude of
    the simulated source.  The completeness values are higher if the
    simulated point sources are inserted in empty regions of the
    images (see Table~\ref{tbl:completeness}).}
  \label{fig:completeness}
\end{figure}

We estimate the completeness of our catalogs as a function of
magnitude by attempting to detect simulated point sources.  The point
sources are created by extracting a bright, non-saturated star,
scaling it to the desired flux level, and inserting it at random
locations in the central well-exposed regions of our $K$-band images.
We then attempt to detect the stars using the same SExtractor settings
described in \S~\ref{sec:photometry}.  Figure~\ref{fig:completeness}
shows the resulting completeness curves as a function of magnitude.

Because the simulated point sources are placed at random locations,
some fraction will fall on or near enough to other objects that
SExtractor does not properly deblend them.  This effect causes the
plateau in the completeness curves at brighter magnitudes to have some
slope; evidently, this problem is worse at fainter magnitudes.  We
repeated the simulations, this time inserting point sources in
locations that avoid other (real or simulated) objects.  The
completeness is now equal to unity to at least $K=20.5$ in all fields.
Table~\ref{tbl:completeness} shows the $90\%$ and $50\%$ completeness
limits for the deep $K$-selected MUSYC catalogs for both the
``unmasked'' and ``masked'' simulations.

We note that the completeness limits from the unmasked and masked
simulations are similar, reflecting the uncrowded nature of the fields
to our depth in $K$.  The masked simulations may provide a better
estimate of the sensitivity of our images, while the unmasked
(shallower) simulations should be used when assessing the completeness
for actual astronomical objects because such objects do not avoid each
other on the sky.  We also note that the completeness for extended
objects will be lower than for the simulated point sources, and would
in principle be a function of inclination, morphology, and size.

\section{Number Counts}
\label{sec:numbercounts}

\begin{figure}
  \epsscale{1.1}
  \plotone{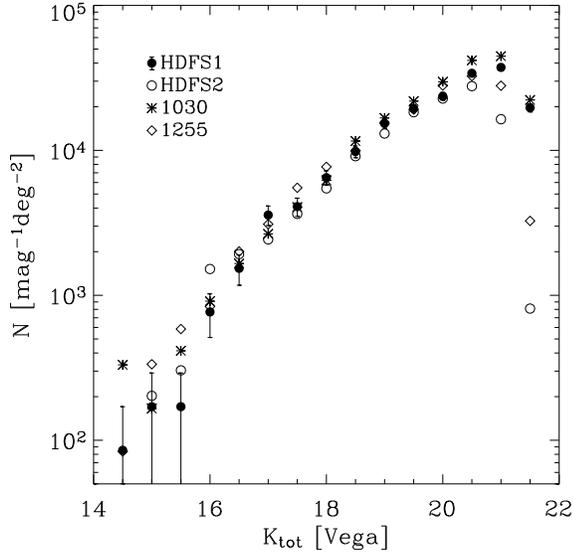}
  \caption{The raw number counts in the four MUSYC fields, excluding
    objects classified as stars.  Poisson errorbars are only shown for
    the HDFS1 field; the other fields have comparable uncertainties.
    No completeness correction has been made.}
  \label{fig:raw_ncounts}
\end{figure}

Figure~\ref{fig:raw_ncounts} shows the surface density of objects as a
function of magnitude, excluding objects classified as stars (\S
\ref{sec:photoz}), for each of our four fields.  No completeness
corrections have been applied.  The number counts are calculated in
$0.5$ magnitude bins using only the image area with $>95\%$ of the
total $K$-band exposure time.  The errorbars assume Poisson
statistics, which should underestimate the true uncertainties because
objects are clustered.  The four fields are generally consistent over
$18 \lesssim K \lesssim 20$, where the S/N is high and differences in
depth do not affect the number counts.  It is noteworthy that the 1030
field has the highest density of objects in this magnitude range; this
field also has the highest density of galaxies with $M >10^{11}
M_\odot$ at $2 < z_{phot} < 3$ of any of the fields studied by
\citet{vandokkum06}, suggesting the possibility of significant galaxy
over-densities in this redshift range.

\begin{figure}
  \epsscale{1.1}
  \plotone{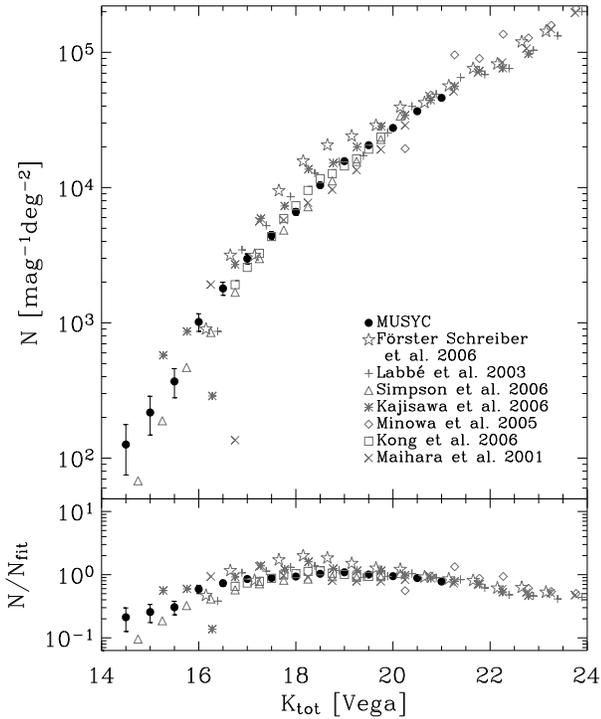}
  \caption{\emph{Top:} The average galaxy number counts from the four
    MUSYC fields, along with a compilation of results from the
    literature.  The MUSYC points have been corrected for
    incompleteness, but only points where this correction is small are
    shown.  \emph{Bottom:} The number counts have been divided by a
    power law with index $\alpha \approx 0.31$, which provides the
    best fit to the MUSYC points over $K \leq 18 \leq 20$.  The
    procedure highlights the increasing logarithmic slope of the
    number counts at $K \lesssim 18$.  The slope may also decrease
    slightly at $K \gtrsim 21$.}
  \label{fig:ave_ncounts}
\end{figure}

The upper panel of Figure~\ref{fig:ave_ncounts} shows the average
number counts of the four fields, along with counts drawn from the
literature.  The completeness corrections described above have been
applied to the MUSYC points.  Note that these completeness values
provide a simplistic correction when dealing with number counts; a
more sophisticated correction would account for the difference between
the measured and intrinsic magnitudes of the artificial sources.  This
can be a significant effect because the rising slope of number counts
means that more sources would scatter to brighter magnitudes than to
fainter magnitudes due to noise fluctuations.  Furthermore, there may
be biases in the photometry of the faintest sources \citep[see
also][]{forster06}.  Neither do we attempt to correct for spurious
detections, which would only be significant at the faintest
magnitudes, or estimate completeness correction for extended sources.
For these reasons we only extend the average number counts to the bin
centered at $K=21$, where the completeness correction begins to become
significant.

The best fit logarithmic slope $d(log N)/dM$ to the MUSYC number
counts is $\alpha \approx 0.31$ over $18 \leq K \leq 20$.  To
illustrate deviations from this power law in different magnitude
ranges, in the lower panel of Figure~\ref{fig:ave_ncounts} we have
divided the observed number counts by this fit.  The deviation from
unity in this panel at brighter magnitudes illustrates a change in the
$d(log N)/dM$ relation.  However, we do not find evidence for a sharp
break in the power law slope at $K \sim 17.5$, as has been reported by
\citet{cristobal03}.  These authors interpret the galaxy number counts
in terms of models of galaxy evolution, and suggest that the break can
be reproduced by models with late star formation in massive galaxies
\citep[ $z\lesssim 2$; see also][]{eliche06}.  There is also some
evidence from deeper surveys that the slope flattens at $K \gtrsim
21$; for instance, the data of \citet{forster06} indicate $\alpha
\approx 0.20$ over $21 \leq K \leq 23$.

\section{Photometric Redshifts and Star Classification}
\label{sec:photoz}

We calculate photometric redshifts using the methods described by
\citet{rudnick01,rudnick03}.  Briefly, non-negative linear
combinations of galaxy templates are fit to the observed spectral
energy distributions.  The templates include the four empirical
templates of \citet{coleman80} and the two empirical starburst
templates of \citet{kinney96}, all of which have been extended into
the UV and NIR using models.  We also include $10\textrm{Myr}$ and
$1\textrm{Gyr}$ old single stellar population templates generated with
the \citet{bruzual03} models, as the empirical templates are derived
from low-redshift galaxies and do not adequately describe some of the
high-redshift galaxies in the MUSYC catalogs.  We do not allow for
additional reddening in the models.  The photometric redshift
uncertainties are calculated using Monte Carlo simulations, in which
the observed fluxes are varied within the photometric uncertainties.
We corrected for Galactic extinction when calculating photometric
redshifts \citep{schlegel98}.

\begin{figure}
  \epsscale{1.1}
  \plotone{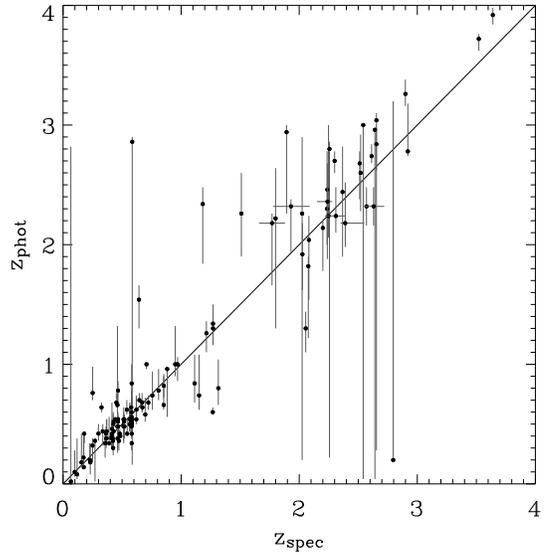}
  \caption{Comparison of photometric and spectroscopic redshifts.
    Errorbars represent the 68\% confidence intervals deteremined from
    Monte Carlo simulations.  The median $|\Delta z| / (1+z)$ is
    0.044.  As described in the text the spectroscopic redshifts for
    six galaxies are shown with some uncertainty, as they are derived
    from fits to the stellar continuum rather than direct observations
    of emission/absorption lines.}
  \label{fig:zphot_comp}
\end{figure}

We have compiled a list of spectroscopic redshifts from the ongoing
MUSYC spectroscopic program \citep[P.~Lira et al.~2007, in
preparation;][]{kriek06b,kriek06c} and the NASA Extragalactic Database.
Figure~\ref{fig:zphot_comp} compares the spectroscopic and photometric
redshifts.  The six spectroscopic redshifts taken from
\citet{kriek06b} are shown with some uncertainty, as they are derived
from fits to the stellar continuum rather than direct measurements of
emission lines.  These galaxies have prominent Balmer/4000\AA\ breaks
that strongly constrain the fits, so there is little chance of
catastrophic failure in the redshift estimates.  In all there are 130
spectroscopic redshifts, 35 of which are at $z>1.5$.

The median of $|\Delta z| / (1+z)$ is 0.044 (0.07 at $z_{spec} >
1.5$).  Catastrophic outliers, which we define as those objects with
$|\Delta z| / (1+z)$ greater than five times the median, comprise $\sim
8\%$ of the sample.  

We find that the SExtractor CLASS\_STAR parameter does not give a
reliable stellar classification for all objects for our $K$-selected
catalogs; some objects with a high stellarity parameter show obvious
extended profiles in the optical bands.  Rather than relying on
SExtractor to identify stars, we fit the NextGen stellar atmosphere
models \citep{hauschildt99} to all objects in the MUSYC catalogs.
Stars are identified as those objects that are better fit by stellar
atmosphere models than by galaxy templates, and which do not have
clearly extended profiles in either the NIR or optical bands.

\section{Analysis of Color Selection Techniques}

In this section we use the MUSYC data and photometric redshifts to
shed light on the relation between galaxies that are selected with the
BM/BX/LBG criteria \citep{steidel03,adelberger04}, the DRG criterion
\citep{franx03}, and the BzK criteria \citep{daddi04}.  These
selection techniques were designed with different types of galaxies in
mind, and for different sets of observational constraints, so it is
not immediately clear how they should relate to each other.  First, we
give a brief discussion of these techniques and describe how we
implement them.  Then we show the locations of galaxies selected using
these criteria on the $BzK$ and $J-K$ diagnostic diagrams, as has been
done previously by \citet{reddy05} for a smaller spectroscopic
sample.  Finally, we present photometric redshift distributions and
rest-frame optical colors for these these galaxies.

The galaxies discussed in this section are drawn from the four deep
MUSYC fields.  We select galaxies with $K \leq 21$ and require a
minimum $K$-band exposure time weight of $0.6$.

\subsection{The color selection criteria}
\label{sec:color_criteria}

\emph{Lyman break and BM/BX galaxies}: The classical ``$U$-dropout''
technique has proven very effective at identifying the so-called Lyman
break galaxies (LBGs) at $z \sim 3$.  \citet{adelberger04} introduced
the BM and BX selection criteria, which are designed to select
galaxies with similar SEDs to the LBGs, but at redshifts $z \sim 1.7$
and $z \sim 2.3$, respectively.  These galaxies are selected to be
blue in the rest-frame UV.  Steidel and collaborators also apply an
${\Rs_{\rm{AB}}} < 25.5$ limit to ensure a reasonable degree of
photometric and spectroscopic completeness.  Because these galaxies
are comparatively bright in the optical and frequently have detectable
emission/absorption lines, they are well-suited to spectroscopic
follow-up \citep[e.g.][]{steidel03}.  However, the BM/BX/LBG
techniques may miss very dusty star-forming galaxies or galaxies with
little ongoing star formation.

The specific color criteria used by \citet{steidel03} and
\citet{adelberger04} are based on $U_nG{\Rs}$ colors, but these
filters were not used as part of MUSYC.  Rather than develop analogous
criteria using our filter set, we calculated synthetic $U_nG{\Rs}$
colors from the best-fitting models \citep[see
also][]{vandokkum04,daddi04,vandokkum06}.  These models were generated
using the \citet{bruzual03} population synthesis code.  In order to
provide the best fits we allowed for a range of star formation
histories and extinction values, but the redshifts were fixed at the
values determined in \S \ref{sec:photoz}.

In the discussion that follows, we draw the distinction between
BM/BX/LBGs that meet the ${\Rs_{\rm{AB}}} < 25.5$ limit and those that
do not.  We also recall that these galaxies are usually selected in the
optical rather than in $K$, so our conclusions may not directly apply to
the wider population of BM/BX/LBGs.

\emph{Distant red galaxies}: The observed $J-K$ color probes the
rest-frame optical at $z \sim 2-4$.  Stars and galaxies at lower
redshifts tend to have blue $J-K$ colors, so the single $J-K>2.3$
criterion for distant red galaxies (DRGs) should primarily select red
galaxies at $z \gtrsim 2$ \citep{franx03,vandokkum03}.  DRGs are
selected in $K$, rather than in the optical, which is closer to a mass
selection than a selection on unobscured star formation.  Because DRGs
are not expected to include galaxies with blue colors or low stellar
masses, they may be complementary to BM/BX/LBGs.  The relative depths
of the MUSYC $JK$ images are well-matched for the identification of
these red galaxies.

\emph{BzK galaxies}: \citet{daddi04} use two sets of color criteria
involving the $B$, $z$, and $K$ bands to select star forming galaxies
at $1.4 \lesssim z \lesssim 2.5$ (hereafter, sBzK galaxies) and
passive galaxies over the same redshift range (hereafter, pBzK
galaxies).  Together, these selection techniques are designed to
identify a nearly complete sample of galaxies in the targeted redshift
range.  The BzK criteria were designed to be used on $K$-selected
objects, so they may not include low-mass galaxies, even when bright
in the rest-frame UV.

The $B$, $z$, and $K_s$ filters used by \citet{daddi04} to define the
BzK color cuts are very similar to the filters used by MUSYC, with the
most significant difference being that the MUSYC $K$ filters extend
slightly further to the blue; this introduces a small $300$\AA\ shift
in effective wavelength.  We make no effort to account for differences
between the filter sets.  Many of the redder galaxies are very faint
or undetected in either $B$ or $z$; for galaxies that are $< 3\sigma$
detections, we use synthetic magnitudes that are calculated directly
from the best-fitting galaxy templates.  The most significant effect
of this procedure is to place objects that are undetected in either
$B$ or $z$ in a reasonable location on the BzK diagram.  

The use of synthetic $B$, $z$ magnitudes for some galaxies is
necessitated by our limited depth.  In fact, the synthetic $B$
magnitudes suggest that many of our reddest $K$-selected galaxies are
so faint in $B$ that they could not be observed in even the deepest
astronomical images.  The same is not true for $z$: the synthetic $z$
magnitudes suggest that the reddest galaxies in our catalogs would
still be detected at $\gtrsim 2\sigma$ in the study recently performed
by \citet{kong06}, which was optimized for the use of the BzK
selection technique.  We note that, to reach this depth, those authors
observed with an 8m telescope.  It follows that, if our $B$ and
$z$ images reached the depth acheived by \citet{kong06}, a
significant number of galaxies would have lower limits in $B-z$.  As
this color is necessary to distinguish between sBzKs and pBzKs (see
Fig.~\ref{fig:bzk}), it may not always be possible to make this
distinction even with very deep optical images; it will, however, be
clear that the galaxies are either sBzKs or pBzKs.  For the sake of
simplicity in the discussion that follows, we assume that the
distinction between sBzKs and pBzKs can always be made.

\subsection{Observed Colors}
\subsubsection{The BzK color-color diagram}

\begin{figure*}
  \plotone{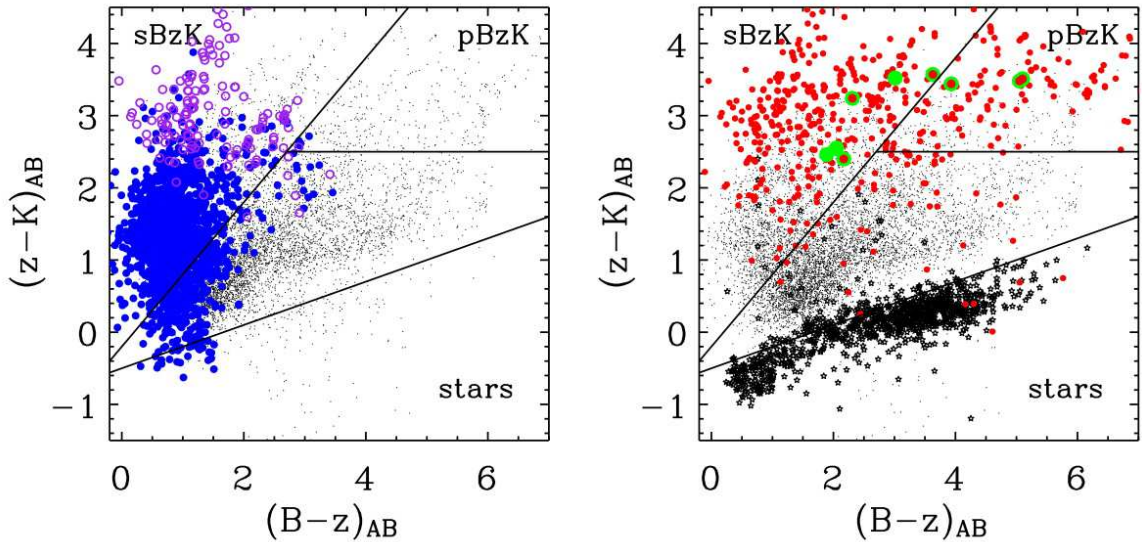}
  \caption{The BzK color-color diagram used by \citet{daddi04} to
    separate star forming galaxies at $z \gtrsim 1.4$, passive
    galaxies at $z \gtrsim 1.4$, and stars.  \emph{Left}: The filled
    blue circles are selected using the BM, BX, or LBG criteria, while
    the open purple circles are galaxies that meet these color
    criteria but are fainter than the typical limit of
    ${\Rs_{\rm{AB}}}<25.5$.  Small black points are the remaining
    MUSYC galaxies.  \emph{Right}: The filled red circles are DRGs.
    The larger filled green circles are the nearly passive galaxies
    from \citet{kriek06b}.  The black stars mark the objects
    identified as stars using our SED-fitting and morphological
    criteria.}
  \label{fig:bzk}
\end{figure*}

Figure~\ref{fig:bzk} shows the $BzK$ color-color diagram, marking the
regions that separate the pBzKs, sBzKs, and stars.  The left panel
highlights the location of BM/BX/LBGs on this diagram.  The majority
of these galaxies fall within the sBzK region, with $\sim 20\%$
falling outside (\citealt{reddy05} find a similar result).  This is
partially due to photometric errors, which can be of significant
concern for galaxies lying near the bottom of the sBzK selection
region.  Since the sBzK selection window was designed to isolate
galaxies at $z \gtrsim 1.4$, it is not a surprise that our photometric
redshifts suggest that $\sim 90\%$ of the BM/BX/LBGs that fall outside
this window are at $z<1.4$.  Roughly half lie at $1<z<1.4$ and half
are the BM/BX/LBG interlopers at $z<1$.  Additionally, some of the
BM/BX/LBGs lie in the region of the $BzK$ color-color diagram that is
supposed to isolate stars \citep[see also][Fig.~12]{reddy05}.

While the sBzK selection appears to identify most of the BM/BX/LBGs at
$z>1.4$, these optical selection techniques are less efficient at
identifying sBzKs: roughly $35\%$ of the sBzKs at our limit of $K<21$
are not selected by any of the optical criteria.  This quantity
increases to $\sim 60\%$ at $K<20$. These values are in rough
agreement with the results from the smaller spectroscopic samples of
\citet{daddi04} (see their Fig.~13) and \citet{reddy05}, and reflect
that fact that a subset of sBzKs have red rest-frame UV colors.
Finally, there is virtually no overlap between the BM/BX/LBGs and
pBzKs.

The right panel of Figure~\ref{fig:bzk} shows the DRGs on the BzK
diagram.  Most of the DRGs lie in either the sBzK region or pBzK
region.  We note that many of the DRGs are very faint in the optical,
so very deep $B$ and $z$ imaging would be necessary to accurately
select these galaxies using the BzK techniques (see
\ref{sec:color_criteria}).  Even so, as shown by \citet{franx03} many
red galaxies escape detection in even the deepest optical images.

The right panel of Figure~\ref{fig:bzk} also shows the locations of 9
galaxies with strongly suppressed star formation, taken from the
sample of \citet{kriek06b}.  The evidence for low ongoing star
formation comes from the lack of detectable rest-frame optical
emission lines in deep NIR spectroscopy, and from stellar population
synthesis modeling which indicates star formation rates of order $\sim
1 M_\odot/yr$.  Despite their nearly passive nature,
Figure~\ref{fig:bzk} shows that only 3 of these 9 galaxies are
classified as pBzKs--with another galaxy very close to the pBzK
selection window--and the remaining galaxies are sBzKs.  It is
possible that this reflects a redshift-dependence in the effectiveness
of the pBzK criterion, even within the range of redshifts for which
this criterion was designed: of the 9 nearly passive galaxies that lie
within or very near to the pBzK region, 4 are at $2.4 \lesssim z \lesssim
2.6$, whereas the others are at $2.0 \lesssim z \lesssim 2.4$.  We
also note that the stellar population modeling performed by
\citet{kriek06b} suggests that the former 4 galaxies may have the
oldest relative ages.  They find $\rm{age} / \tau \gtrsim 25$ for
these 4 galaxies, where $\tau$ is the star formation $e$-folding
timescale, and the remaining 5 have $\rm{age} / \tau \lesssim 25$.
Thus it could also be that pBzK criterion only works well for the oldest
subset of nearly passive galaxies.

\subsubsection{The $J-K$ versus $K$ color-magnitude diagram}

\begin{figure*}
  \plotone{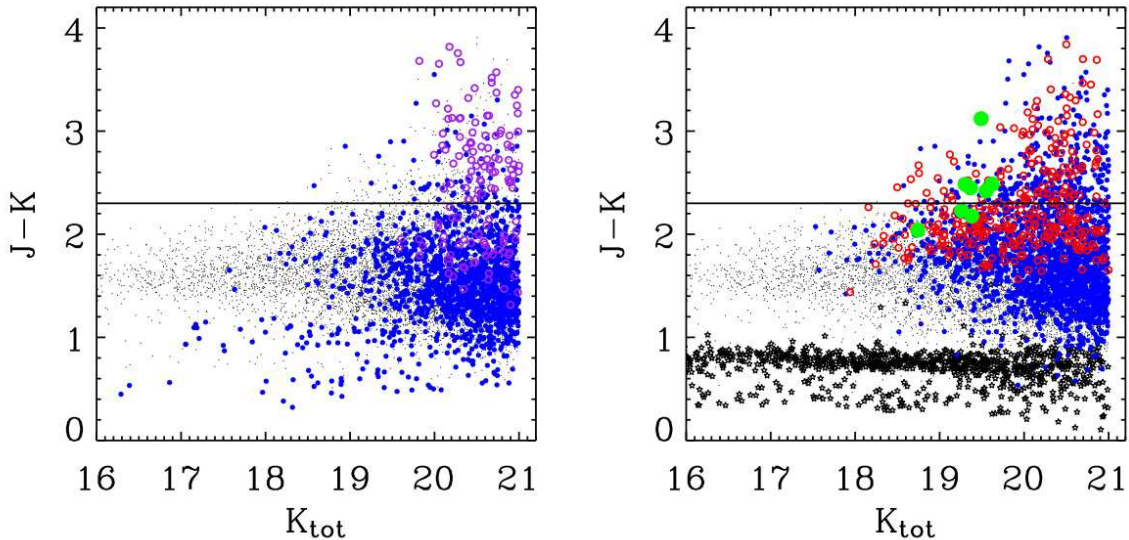}
  \caption{$J-K$ versus $K$.  The lines at $J-K = 2.3$ illustrate the
    DRG limit.  \emph{Left}: The filled blue circles are selected
    using the BM, BX, or LBG criteria, while the open purple circles
    are galaxies that meet these color criteria but are fainter than
    the typical limit of ${\Rs_{\rm{AB}}} = 25.5$.  Small black points
    are the remaining galaxies.  A slight majority of BM/BX/LBGs which
    are also classified as DRGs are fainter than this limit.
    \emph{Right}: Filled blue circles are sBzKs, and open red circles
    are pBzKs.  The larger filled green circles are the nearly passive
    galaxies from \citet{kriek06b}.  Two of the three passive galaxies
    that are not classified as DRGs are within $\sim 1\sigma$ of the
    DRG limit.  The black stars mark the objects identified as stars
    using our SED-fitting and morphological criteria.}
  \label{fig:jk}
\end{figure*}

Figure~\ref{fig:jk} shows the $J-K$ versus $K$ diagram, along with a
line at $J-K = 2.3$ which marks the limit for DRGs.  The left panel
shows that only $\sim 10\%$ of BM/BX/LBGs are classified as DRGs.
Conversely, $\sim 32\%$ of DRGs satisfy one of the BM/BX/LBG criteria.
Of the galaxies that meet both sets of criteria, $\sim 55\%$ are
fainter than the ${\Rs_{\rm{AB}}} = 25.5$ limit used in many
ground-based samples of optically-selected galaxies.\footnote{These
  results do not change significantly if the $z \sim 1.4$ BM galaxies
  are removed from the analysis.}  Thus the overlap between DRGs and
BM/BX/LBGs is small, and has a significant contribution from galaxies
that are too faint to be included in typical optical surveys
(i.e.~they have ${\Rs_{\rm{AB}}} > 25.5$) \citep[see also][]{reddy05,
  vandokkum06}.  

The right panel shows that sBzKs and pBzKs span a range in $J-K$
color.  This panel also shows the location of the nearly passive
galaxies described by \citet{kriek06b}.  The DRG criterion identifies
$6/9$ of these galaxies.  This incompleteness may be due to
photometric uncertainties, as $2/3$ of the remaining galaxies are
within $\sim 1\sigma$ of the DRG limit.  As shown above, only $3/9$
are selected by the pBzK criterion.

\subsection{Photometric redshift distribution}

\begin{figure}
  \epsscale{1.1}
  \plotone{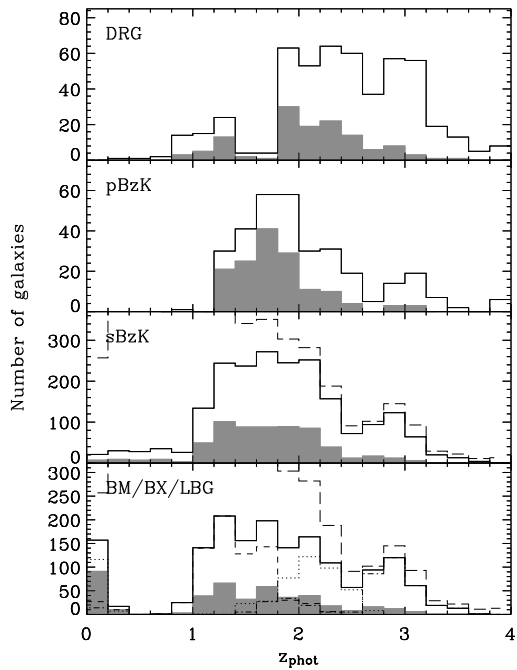}
  \caption{The photometric redshift distribution of $K<21$ galaxies
    selected according to different selection criteria.  The filled
    histograms are for galaxies with $K<20$.  The dashed, dotted, and
    dot-dashed histograms on the bottom panel are for the BM, BX, and
    LBG criteria, respectively.  The upper dashed lines in the lower
    panels show the redshift distribution of all objects in the MUSYC
    catalogs, illustrating that the sBzK and BM/BX/LBG criteria select
    most of the galaxies in the higher redshift bins.}
  \label{fig:redshift_hist}
\end{figure}

Figure~\ref{fig:redshift_hist} shows the photometric redshift
distributions for each of the color criteria discussed above.  The
distribution of DRGs shows a prominent gap in the region $z \simeq
1.4-1.8$.  It is not clear whether this is an artifact in our
photometric redshift or a real feature.  Defining interlopers as those
objects at $z<1.8$, the interloper fraction is $\sim 15\%$ at $K<21$,
$\sim 20\%$ at $K<20$, and $\sim 50\%$ at $K<19$.  

The photometric redshift distribution of pBzKs is shown in the second
panel of Figure~\ref{fig:redshift_hist}.  The photometric redshifts
indicate that this selection technique is indeed effective at
isolating galaxies at $z>1.4$.  The interlopers lie primarily at
$1.2<z<1.4$, with very few at lower redshift.  As shown in the third
panel, the sBzK galaxies have an approximately similar redshift
distribution but with a substantial number reaching down to $z \sim
1-1.2$.  The relatively minor differences in photometric redshift
distributions suggest that these techniques may be used to select
complementary samples of galaxies, although the difference in the
number of interlopers should be taken into account: $10\%$ of the
pBzKs and $25\%$ of the sBzKs lie at $z<1.4$ (these numbers increase
to $14\%$ and $30\%$, respectively, for $K<20$, and to $22\%$ and
$30\%$ for $K<19$).  Also, as shown above, galaxies with quenched star
formation may be classified as sBzKs.  For these reasons, it appears
that precise comparisons of the global properties of star forming and
passive galaxies (e.g. number density, stellar mass, or clustering)
using only $BzK$ photometry may be of limited usefulness
\citep[see][]{kong06}.

While the BzK criteria were designed to select galaxies at $1.4 \leq z
\leq 2.5$, our photometric redshifts suggest that they identify a
fairly complete sample of galaxies at larger redshifts, up to $z \sim
3.5$.  Together, these criteria are very effective at selecting high
redshift galaxies: $\sim 93\%$ of the galaxies in our catalogs at $1.4
\leq z \leq 3.5$ satisfy either sBzK or pBzK criteria.  This can be
seen as either a benefit or a drawback to these selection criteria; a
benefit because only three bands are needed to isolate a large number
of high redshift galaxies; a drawback because global properties of
BzK-selected galaxies will be an average over a wide range in
redshift, obscuring the evolving nature of galaxies at an epoch where
such evolution is expected to be rapid.

It is also noteworthy that pBzKs have the highest fraction of galaxies
at $K<20$ of any of the galaxy populations discussed here.  The sBzKs in
particular have a smaller fraction of galaxies this bright at the same
redshifts.  This may suggest that pBzKs have a flatter luminosity
function in the rest-frame optical than do the sBzKs, although a
detailed discussion of this point is clearly beyond the scope of this
paper.  A similar conclusion was reached about the luminosity functions
of $z \sim 2.5$ red and blue galaxies by \citet{marchesini06}.

The photometric redshift distributions of BM/BX/LBGs are shown in the
bottom panel of Figure~\ref{fig:redshift_hist}.  Excluding the $z<0.4$
interlopers, the mean photometric redshifts for the three galaxy
samples are $1.4$, $2.1$, and $2.8$, respectively.  These values are
in good agreement with the spectroscopic values of
\citep{steidel03,steidel04}, although our $K$-band selection certainly
means that our sample will have different properties than the bulk of
typical $\Rs$-selected samples
\citep{shapley04,adelberger05b,reddy05}.  Similarly, the $\sim 13\%$
interloper fraction for BM/BX galaxies is in excellent agreement with
the spectroscopic sample of \citet{reddy05} (see their Table 3).  A
strength of the BM/BX/LBG selections is that each identifies galaxies
over a comparatively narrow redshift range, so comparisons of the
properties of galaxies selected by each of these techniques provide
meaningful constraints on evolution \citep{adelberger05a}.

\begin{figure*}
  \plotone{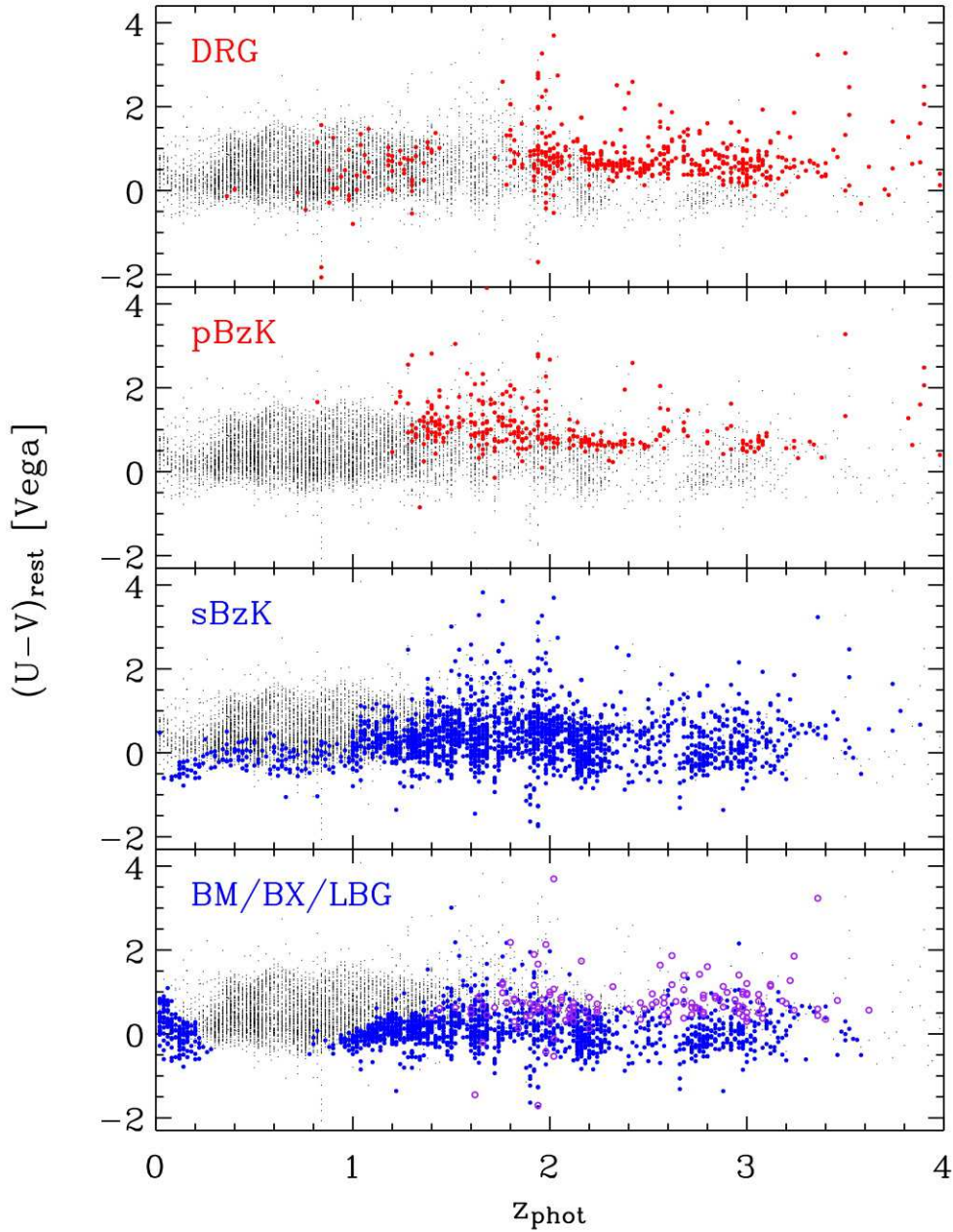}
  \caption{Rest-frame U-V color vs. redshift.  The filled circles mark
    the different galaxy populations discussed in this paper, and the
    small black points mark the remaining MUSYC galaxies.  The open
    purple circles in the bottom panel are for galaxies that meet the
    BM/BX/LBG color criteria, but are fainter than the typical limit
    of ${\Rs_{\rm{AB}}}=25.5$}
  \label{fig:uv_z}
\end{figure*}

\subsection{Rest-frame optical colors}

Next we investigate the rest-frame optical U-V color of galaxies as a
function of redshift.  The rest-frame colors were calculated by
interpolating between the observed bands using the best-fitting
templates as a guide \citep[see][]{rudnick03}.  The results are shown
in Figure~\ref{fig:uv_z}.  As the $J-K$ color probes the rest-frame
optical at $z \gtrsim 2$, it is no surprise that the DRGs at these
redshifts tend to be red.  The ``interlopers'' at $z<1.8$ have a wider
range of colors.  Compared to the pBzK selection technique, the DRG
technique does appear to select a larger number of red galaxies at $z
> 2$, identifying $\sim 75\%$ of all galaxies with $(U-V)_{rest} >
0.5$ (the remaining $\sim 25\%$ are nearly all at $2<z<2.3$).  For
comparison the pBzK technique only identifies $\sim 30\%$ of these
galaxies, but it extends to lower redshifts and has very few
interlopers.  It should also be emphasized that the pBzK technique was
designed to select only passive galaxies, while the DRG technique
selects both passive and actively star forming galaxies
\citep[e.g.][]{labbe05,kriek06b}; it is not currently known if there
is significant contamination of pBzK samples by star forming galaxies.

As shown above, the sBzK technique identifies most of the galaxies at
$z > 1.4$.  Interestingly, there is a significant tail of the redshift
distribution at $1 \leq z \leq 1.4$--as well as a more sparsely
populated tail at lower redshifts--that seems to be occupied only by
bluer galaxies.  This redshift-dependent selection effect may skew any
global quantities derived from samples of sBzK galaxies.

The BM/BX/LBG selection criteria identify the bluer galaxies.  The
blue points in the bottom panel of Figure~\ref{fig:uv_z} meet the
${\Rs_{\rm{AB}}}<25.5$ limit used for many optical surveys, while this
magnitude cut has not been imposed for the purple points.  As expected
in a $K$-selected catalog, galaxies that are fainter in $\Rs$ tend to
have redder colors.  It has been noted previously that many high
redshift galaxies do not meet the BM/BX/LBG criteria because they are
too red--not simply because they are too faint in observer's optical
\citep{vandokkum04,daddi04,vandokkum06}.  Interestingly, while $\sim
50\%$ of the galaxies at $z \sim 2$ meet one of the optical selection
criteria, this number increases to $\sim 80\%$ at $z \sim 3$ (see also
the bottom panel of Fig.~\ref{fig:redshift_hist}).  This suggests that
a higher fraction of $z \sim 3$ galaxies have blue continuua and
prominent Lyman breaks, indicative of significant unobscured star
formation.  This apparent evolution is not caused by the Malmquist
bias: applying a uniform cut in absolute $V$-band magnitude--the cut
is chosen to ensures high completeness at $z \sim 3$--the fraction of
galaxies that are selected by any of the BM/BX/LBG criteria is $\sim
35\%$ and $\sim 80\%$ at $z \sim 2$ and $z \sim 3$, respectively.  An
important caveat to this result is that $z \sim 2$ galaxies are
selected using the BM and BX criteria, while $z \sim 3$ galaxies are
selected using the LBG criteria.  It may thus be possible that this
effect is an artifact of differences in the color selection criteria
themselves, rather than evidence of evolution.  However
\citet{brammer06} study the evolution of optically-red galaxies from
$z \sim 3.7$ to $z \sim 2.4$, and also find evidence for substantial
evolution in the rest-frame ultra-violet slopes in the same sense as
described here.

\section{Summary}
\label{sec:conclusions}

We have presented the deep NIR imaging of the MUSYC survey.  This
consists of four $10\arcmin \times 10\arcmin$ fields, imaged to $J
\sim 22.5$, $H \sim 21.5$, and $K \sim 21$.  We combined these data
with MUSYC optical imaging to produce public $K$-selected catalogs with
uniform $UBVRIzJHK$ photometry. 

Many recent surveys rely on a few observed bands to isolate large
samples of high redshift galaxies.  We use the high-quality,
multi-band photometry from MUSYC to investigate some of the properties
of galaxies that are selected using common selection criteria,
including the distant red galaxies (DRGs), star-forming BzK galaxies
(sBzKs), passive BzK galaxies (pBzKs), as well as the Lyman break
galaxies (LBGs) and the similar BM/BX galaxies at somewhat lower
redshifts.

DRGs have a wide distribution of photometric redshifts.  The DRGs at
$z \gtrsim 1.8$ have red rest-frame optical colors, while the
interlopers at lower redshift have a range of optical colors.  The
interlopers account for $\sim 15\%$ of the DRGs at $K < 21$, but this
number increases to $\sim 50\%$ at $K < 19$.  In comparison, the
BM/BX/LBGs each have comparatively narrow redshift windows.  If an
${\Rs_{\rm{AB}}}<25.5$ magnitude limit is applied, which is a typical
limit for ground-based surveys of optically-selected galaxies, then
the BM/BX/LBGs tend to have blue rest-frame optical colors.  If this
limit is not applied then these selection criteria also identify many
galaxies with redder colors, including some that are DRGs.
Interestingly, the combined BM/BX/LBG selection criteria identify a
significantly higher fraction of the $K$-selected galaxies at $z \sim
3$ than at $z \sim 2$; this may suggest an evolution in the rest-frame
UV properties of red galaxies, as has already been found by
\citet{brammer06}.  Current samples of DRGs and BM/BX/LBGs are largely
orthogonal, but the overlap is expected to increase for optical
surveys that reach limits significantly deeper than ${\Rs_{\rm{AB}}} =
25.5$.  Together, the DRG and BM/BX/LBG criteria select $\sim 90\%$ of
the galaxies with $2 < z < 3.5$ in our sample.

The sBzK and pBzK criteria select galaxies over a very wide range in
redshift.  While the sBzK criterion tends to select only galaxies with
bluer rest-frame optical colors at $1 \leq z \leq 1.4$, it also
selects many of the reddest galaxies at higher redshifts.  Because the
sBzK criterion selects many different types of galaxies over a large
range in redshift, it may be thought of as a coarse but effective
photometric redshift technique.  Together, the sBzK and pBzK criteria
detect $\sim 93\%$ of the galaxies at $1.4 \leq z \leq 3.5$ to our
limit of $K=21$.  A larger fraction of the pBzKs than sBzKs are bright
in $K$, which may be indicative of a difference in luminosity
functions.  Finally, the usefulness of the BzK criteria may be
limited for the reddest galaxies, which may be undetected in even the
deepest optical images.

A thorough understanding of the range of properties spanned by
galaxies selected according to specific color criteria requires
multiwavelength observations and a large sample of spectroscopic
redshifts.  The use of photometric redshifts is a significant
limitation of the analysis presented in this paper.  Unfortunately, a
large number of spectroscopic redshifts for an \emph{unbiased} sample
of the galaxies detected by current deep surveys, including those with
very red colors, is exceedingly difficult to obtain.  Progress in the
study of galaxy properties as a function of redshift will continue to
rely heavily on color selection techniques, both to define galaxy
samples and to identify targets for spectroscopic follow-up.  An
awareness of the limitations of such techniques is advisable.

\acknowledgements

We are grateful to members of the MUSYC collaboration for enabling
this research.  We thank Ivo Labb\'{e} for several illuminating
discussions, Natascha F\"orster Schreiber for providing information
about data reduction and NIR surveys, and the CTIO staff for their
assistance with ISPI observations.  This publication makes use of data
products from the Two Micron All Sky Survey, which is a joint project
of the University of Massachusetts and the Infrared Processing and
Analysis Center/California Institute of Technology, funded by the
National Aeronautics and Space Administration and the National Science
Foundation.  MUSYC has benefited from the support of Fundaci\'{o}n
Andes.  We acknowledge support from NSF CAREER AST-0449678. DM is
supported by NASA LTSA NNG04GE12G. EG is supported by NSF Fellowship
AST-0201667.  PL is supported by Fondecyt Project \#1040719.

Facilities: \facility{CTIO.VBT(ISPI)}

\begin{deluxetable}{cccccc}
\tablecolumns{6}
\tablecaption{Deep NIR MUSYC Fields}
\tablehead{ \colhead{Field} & \colhead{RA (J2000.0)} & \colhead{Dec (J2000.0)} 
& \colhead{$E(B-V)$} & \colhead{Filter} & \colhead{Exposure time (hr)} }
\startdata
HDFS1 & 22:33:12.5 & -60:36:40 & 0.027 & $J$ & 18.0 \\
      &            &           &       & $H$ & 8.3 \\
      &            &           &       & $K'$ & 7.0 \\
HDFS2 & 22:31:56.6 & -60:36:46 & 0.026 & $J$ & 10.3 \\
      &            &           &       & $H$ & 5.0 \\
      &            &           &       & $K_s$ & 8.5 \\
1030 & 10:30:30.4 & +5:25:00 & 0.024 & $J$ & 11.3 \\
         &            &          &       & $H$ & 9.9 \\
         &            &          &       & $K'$ & 10.4 \\
1255 & 12:55:20.6 & +1:07:49 & 0.015 & $J$ & 10.7 \\
         &            &          &       & $H$ & 6.9 \\
         &            &          &       & $K_s$ & 11.3 \\

\enddata
\label{tbl:field_chars}
\end{deluxetable}

\begin{deluxetable}{cccc}
\tablecolumns{4}
\tablecaption{AB conversions}
\tablehead{ \colhead{Field} & \colhead{filter} 
& \colhead{$\lambda_{\textrm{eff}}$ [\AA]} 
& \colhead{AB conversion \tablenotemark{a}}}
\startdata
HDFS1, 1030 & $J$ & 12461 & 0.93 \\
            & $H$ & 16306 & 1.38 \\
            & $K'$ & 21337 & 1.86 \\
HDFS2, 1255 & $J$ & 12470 & 0.94 \\
           & $H$ & 16366 & 1.40 \\
           & $K_s$ & 21537 & 1.88 \\

\enddata
\tablenotetext{a}{Defined such that $m_{AB} = m_{Vega} + conversion$}
\label{tbl:AB_conversions}
\end{deluxetable}

\begin{deluxetable}{ccccc}
\tablecolumns{5}
\tablecaption{Characteristics of final images}
\tablehead{ \colhead{Field} & \colhead{Filter} 
& \colhead{Zeropoint (Vega)} & \colhead{FWHM} & \colhead{$5\sigma$ depth} }
\startdata
HDFS1 & $J$ & $22.37 \pm 0.02$ & 0.96 & 22.9 \\
      & $H$ & $22.495 \pm .015$ & 0.96 & 21.8 \\
      & $K'$ & $22.274 \pm .02$ & 0.96 & 21.1 \\
HDFS2 & $J$ & $22.127 \pm .025$ & 1.05 & 22.5 \\
      & $H$ & $22.522 \pm .023$ & 1.03 & 21.4 \\
      & $K_s$ & $22.094 \pm .022$ & 1.00 & 20.8 \\
1030 & $J$ & $22.356 \pm .02$ & 0.97 & 22.5 \\
         & $H$ & $22.545 \pm.015$ & 0.93 & 21.8 \\
         & $K'$ & $22.23 \pm .03$ & 0.90 & 21.3 \\
1255 & $J$ & $21.959 \pm .012$ & 1.00 & 22.6 \\
         & $H$ & $22.463 \pm .017$ & 0.92 & 21.6 \\
         & $K_s$ & $22.115 \pm .020$ & 0.93 & 21.0 \\

\enddata
\label{tbl:image_chars}
\end{deluxetable}

\begin{deluxetable}{ccccc}
\tablecolumns{5}
\tablecaption{Point-source completeness limits for $K$-selected catalogs}
\tablehead{ 
  \colhead{Field} & 
  \multicolumn{2}{c}{\underline{Masking sources\,}} & 
  \multicolumn{2}{c}{\underline{Entire image\,}} \\
  \colhead{} &
  \colhead{90\% limit} & 
  \colhead{50\% limit} & 
  \colhead{90\% limit} &
  \colhead{50\% limit} }
    \startdata
    HDFS1 & 21.18 & 21.50 & 20.78 & 21.44 \\
    HDFS2 & 20.79 & 21.10 & 20.55 & 21.04 \\
    1030  & 21.37 & 21.68 & 21.15 & 21.59 \\
    1255  & 20.98 & 21.30 & 20.93 & 21.27 \\
    \enddata
    \label{tbl:completeness}
\end{deluxetable}

\end{document}